\documentclass[aip,amsmath,amssymb,reprint, twocolumn]{revtex4-1}

\usepackage[mathlines]{lineno}
\usepackage[utf8]{inputenc}
\usepackage[T1]{fontenc}
\usepackage{mathptmx}

\usepackage{lipsum}
\usepackage{listings}
\usepackage{xcolor}
\usepackage{siunitx}
\definecolor{codegreen}{rgb}{0,0.6,0}
\definecolor{codegray}{rgb}{0.5,0.5,0.5}
\definecolor{codepurple}{rgb}{0.58,0,0.82}
\definecolor{backcolour}{rgb}{0.95,0.95,0.92}
 
\lstdefinestyle{mystyle}{
    backgroundcolor=\color{backcolour},   
    commentstyle=\color{codegreen},
    keywordstyle=\color{magenta},
    numberstyle=\tiny\color{codegray},
    stringstyle=\color{codepurple},
    basicstyle=\ttfamily\footnotesize,
    breakatwhitespace=false,         
    breaklines=true,                 
    captionpos=b,                    
    keepspaces=true,                 
    numbersep=5pt,                  
    showspaces=false,                
    showstringspaces=false,
    showtabs=false,                  
    tabsize=2
}
 
\usepackage{times,amsmath}
\usepackage{epsfig}
\usepackage{color}
\usepackage{longtable}
\usepackage{ulem}
\usepackage{graphicx}
\usepackage{dcolumn}
\usepackage{bm}
\usepackage{bookmark}
\usepackage{tabularx}
\usepackage{hyperref}
\usepackage{multirow}
\hypersetup{colorlinks=true, citecolor=blue, filecolor=blue, linkcolor=blue, urlcolor=blue}

\begin{document}

\title{Symmetry Relation Database and Its Application to Ferroelectric Materials Discovery}

\author{Qiang Zhu}
\email{qiang.zhu@unlv.edu}
\affiliation{Department of Physics and Astronomy, University of Nevada, Las Vegas, NV 89154, USA}

\author{Byungkyun Kang}
\affiliation{Department of Physics and Astronomy, University of Nevada, Las Vegas, NV 89154, USA}

\author{Kevin Parrish}
\affiliation{Department of Physics and Astronomy, University of Nevada, Las Vegas, NV 89154, USA}

\date{\today}
\begin{abstract}
The ability to understand the atomistic mechanisms that occur in the solid phase transition is of crucial importance in materials research. To investigate the displacive phase transition at the atomic scale, we have implemented a numerical algorithm to automate the detection of the symmetry relations between any two candidate crystal structures. Using this algorithm, we systematically screen all possible polar-nonpolar structure pairs from the entire Materials Project database and establish a database of $\sim$4500 pairs that possess a close symmetry relation. These pairs can be connected through a continuous phase transition with small atomic displacements. From this database, we identify several new ferroelectric materials that have never been reported in the past. In addition to the screening of ferroelectric materials, the symmetry relation database may also be used for other areas, such as material structure prediction and new materials discovery. 
\end{abstract}


\maketitle


\section{INTRODUCTION}
The solid-solid phase transitions refer to the phase change between different crystalline forms of the same compound. Among them, the phase transitions driven by temperature are of note. Though these transitions often involve just minor structural modifications through displacive mechanisms, they can have a profound effect on a variety of the solid's physical properties. According to Landau theory \cite{landau2013statistical}, continuous phase transitions require that there is a group-subgroup relation between the involved crystal structures. Knowing the atomic details that occur during the phase relation plays a pivotal role in materials research. For instance, detecting possible symmetry-related structures is the basis for the identification of polar materials that are likely to have a ferroelectric to paraelectric phase transition at high temperatures \cite{shi2016symmetry, smidt2020automatically}. In addition, there is a great potential to reduce the complexity of the search space in crystal structure prediction \cite{Zhu-CE-2012, QZhu-PRB-2015} by utilizing symmetry relations. 

The group-theoretical presentation of crystal-chemical relationships have been long established. It is possible to present symmetry relations between two crystal structures in a concise manner with a Barnighausen tree \cite{wondratschek2006symmetry}. 
To investigate a single phase transition, an experienced researcher often resorts to the text book or online server to derive the symmetry relation by hand. Despite the fact that we are entering the data-driven era in materials research, most of previous efforts in phase transition were limited to case studies. To our knowledge, there is no attempt to systematically investigate the symmetry relation from the historic materials data, which prevents more effective materials design from the knowledge of crystallographic symmetry.

In this paper, we start by describing the implementation of a numerical algorithm that enables the high-throughput detection relation from any crystallographic data. Using this tool, we search the possible polar/non-polar structure pairs from the entire Materials Project (MP) database and establish a database containing 4514 structure pairs that possess close symmetry relation. In particular, we identify 1234 potential ferroelectric-paraelectric pairs. Lastly, we discuss the possible research directions for utilizing the symmetry relation database as a tool for future materials development.

\begin{figure*}[ht]
\centering
\includegraphics[width=0.95\textwidth]{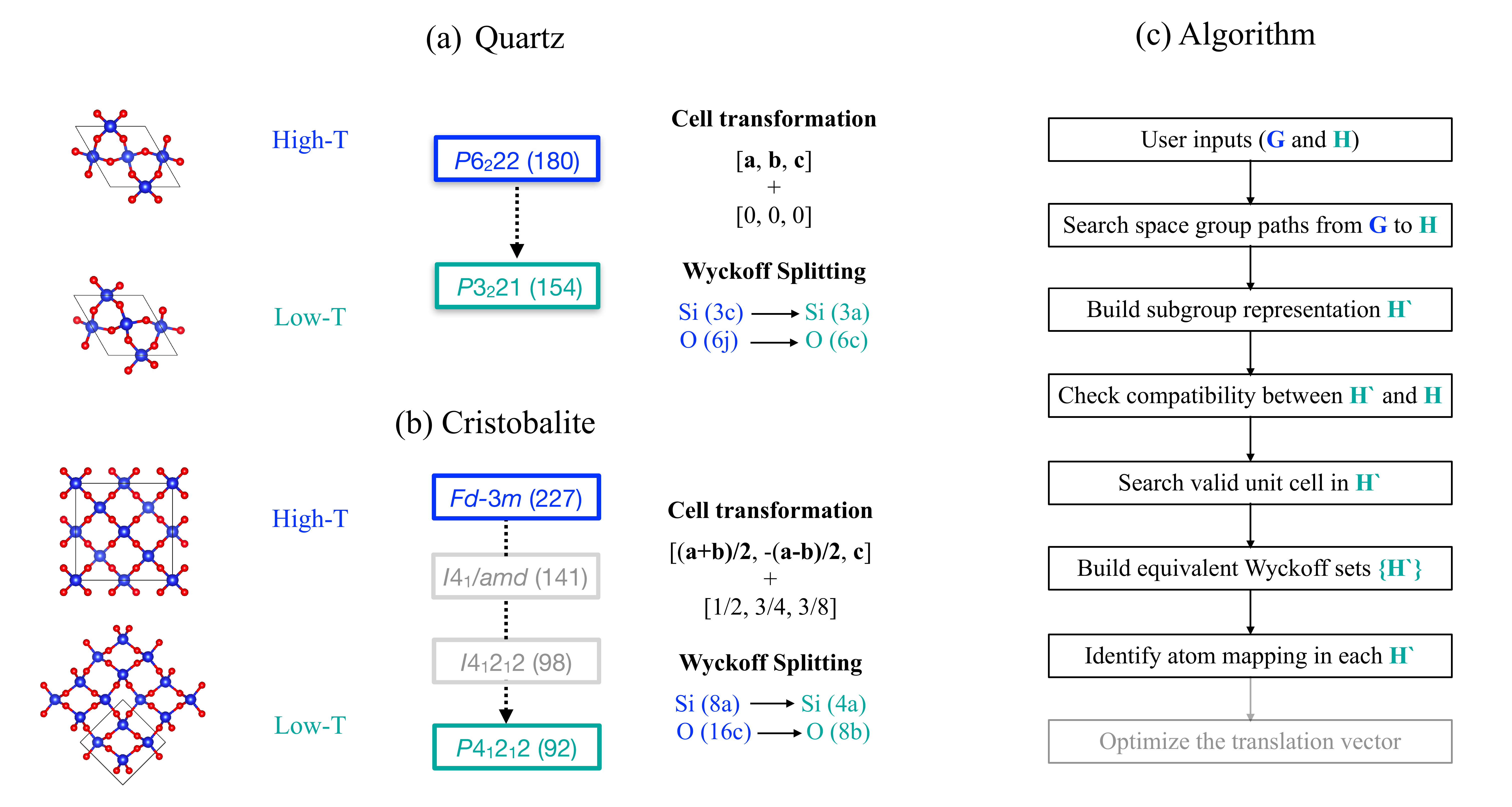}
\caption{The schematic phase transition. (a) and (b) illustrate the mechanism for different silica polymorphs of quartz and cristobalite, respectively. (c) outlines the workflow for the symmetry relation detection algorithm. For quartz, it follows a one step symmetry relation between the high temperature $\beta$ (hexagonal) and low temperature $\alpha$ (trigonal) phases. 
On the other hand, the transition between two phases of cristobalite requires multiple steps of group-subgroup transitions.}
\label{fig1}
\end{figure*}

\section{The symmetry relation detection algorithm} \label{sec:method}
Relations between crystal structures can often be expressed by group–subgroup relations. A subgroup $H$ of a space group $G$ is called a maximal subgroup of $G$ if there is no proper subgroup $M$ in between, i.e., $G \rightarrow M \rightarrow H$. There exist two kinds of $H$ subgroups. The \textit{translationengleiche} ($t$) subgroups keeps all translational symmetries but belong to a lower crystal class, while the \textit{klassengleiche} ($k$) subgroups have fewer translations but are of the same crystal class. For a given space group, its maximal t and k subgroups are well defined and can be found in the crystallography book \cite{aroyo2013international}. To complete the transition between a group $G$ and its maximal subgroup $H$, one needs to know the cell transformation matrix, as well as the Wyckoff position (WP) splitting scheme for the given atomic positions. 

If the phase transition follows the immediate symmetry relation ($G \rightarrow H$), one needs to (i) derive the unit cell transformation; (ii) map all WPs in both $G$ and $H$ to ensure that they follow the required splitting relation; and (iii) compute the distortions of unit cell vectors and atomic coordinates. While this procedure for mapping two structures is straightforward for a simple case such as the transition between $\alpha \rightarrow \beta$ phases of silica quartz (see Fig. \ref{fig1}a), many transitions may follow a chain of subgroup relations between two crystals (e.g., $G \rightarrow M_1 \rightarrow M_2 \rightarrow H$, see an example transition between $\alpha \rightarrow \beta$ phases of silica cristobalite in Fig. \ref{fig1}b). In such a case, a more complicated search for possible transition pathways in the large search space connecting $G$ and $H$ is required. When it comes to the low symmetry cases, the mapping of the crystal lattice between transition structures is not unique and there exists multiple choices for arrangements of the cell vectors. When it comes to the mapping of the different WP sites between the low and high-symmetry structures, the atomic coordinates may be continuously shifted to better match the reference states (\textit{i.e.}, the $G$ and $H$ structures). In this case, finding the optimum distortion path is not trivial as well. 

\subsection*{2.1 Numerical implementation}
To address the aforementioned challenges, we have developed the following search algorithm to check if two input crystals follow symmetry relations (see Fig. \ref{fig1}c).

\begin{enumerate}
    \item \textbf{Space group transition path search.} In order to find the path between two arbitrary space groups ($\boldsymbol{G}$ and $\boldsymbol{H}$), we will adopt the depth-first search algorithm for traversing tree data structures. Starting with high symmetry $\boldsymbol{G}$, we will enumerate all possible maximum $t$ and $k$ subgroups ($\boldsymbol{\{H_1\}}$), and then the next layer subgroups $\boldsymbol{\{H_{i+1}\}}$ will be listed from each member in the previous layer $\boldsymbol{\{H_{i}\}}$). The iteration will be terminated once it reaches the maximum number of layers; the branch will stop growing when the target $\boldsymbol{H}$ is found. For $k$-subgroup transitions along this path, there exist an infinite number of ways to generate the new subgroup symmetry within the same point group by replicating the unit cell along a crystallographic axis. To account for such cases that require more intermediate $k$-subgroups, we run the search twice. In the first round, we only examine the subgroup chain sequence with the shortest possible transition sequence between $\boldsymbol{G}$ and $\boldsymbol{H}$. If this search does not lead to any valid solution that can map the corresponding Wyckoff letters between $\boldsymbol{G}$ and $\boldsymbol{H}$, we will explore more pathways by considering extra $k$-type subgroups for each of the navigated transitions in the first round. 
    
    \item \textbf{Compatibility check}. For each given path, we will transform the high symmetry structure into a subgroup setting ($\boldsymbol{H'}$) and check if the resulting WP sites in the subgroup setting are consistent with those in the low symmetry structure. The goal is to ensure that groups of similar atomic species are compatible with the Wyckoff positions of both structures.

    \item \textbf{Find the matched unit cell}. If the target $\boldsymbol{H}$ has a tetragonal symmetry or higher, one just need to directly compare the cell vectors between $\boldsymbol{H'}$  and $\boldsymbol{H}$. When $\boldsymbol{H}$ belongs to the orthorhombic symmetry, the permutation of crystallographic $a, b, c$ axes also need to be considered in order to achieve the best match. For the monoclinic and triclinic cells, one also needs to check other possible unit cell settings such as ($a+b$, $a-b$, $c$) without changing the symmetry operation. Only the structures of $\boldsymbol{H'}$ having the close unit cell within a tolerance of 1.5 \AA~ will be considered for the next step.
    
    \item \textbf{Enumerate all equivalent Wyckoff sets $\boldsymbol{\{H'\}}$}. In a crystal structure, the atoms may be equivalently described by different sets of WPs. For instance, a FCC Cu can be considered as either WP 4a (0, 0, 0) or 4b (1/2, 1/2, 1/2). All different Wyckoff settings can be computed by the Euclidean and the affine normalizers of a given space group \cite{wondratschek2006symmetry}. For each $\boldsymbol{H'}$ with a valid unit cell representation, 
    we will then enumerate all possible equivalent discrete sets of WPs in order to identify additional compatible transition pathways.
    
    \item \textbf{Atom mapping}. For each member in $\boldsymbol{\{H'\}}$, we will navigate all likely matched pairs between WP1 in $\boldsymbol{H'}$ and WP2 in $\boldsymbol{H}$. According to the choice of $\boldsymbol{H}$, there may exist a continuous translation to obtain the equivalent Wyckoff set. If translation is possible, we will perform an overall translation between WP1 and WP2. Finally, we compute the relative atomic displacements in order to match all WP pairs in $\boldsymbol{H'}$ and $\boldsymbol{H}$. 
    
    \item \textbf{Optimization of translation vector (Optional)}. If a continuous translation is allowed, we will also perform a gradient-free optimization to minimize the atomic displacements. Within the framework of Python, this can be conveniently implemented based on the Nelder-Mead algorithm \cite{gao2012implementing} in the SciPy \cite{scipy}.
\end{enumerate}

\begin{figure}[htbp]
\centering
\includegraphics[width=0.4\textwidth]{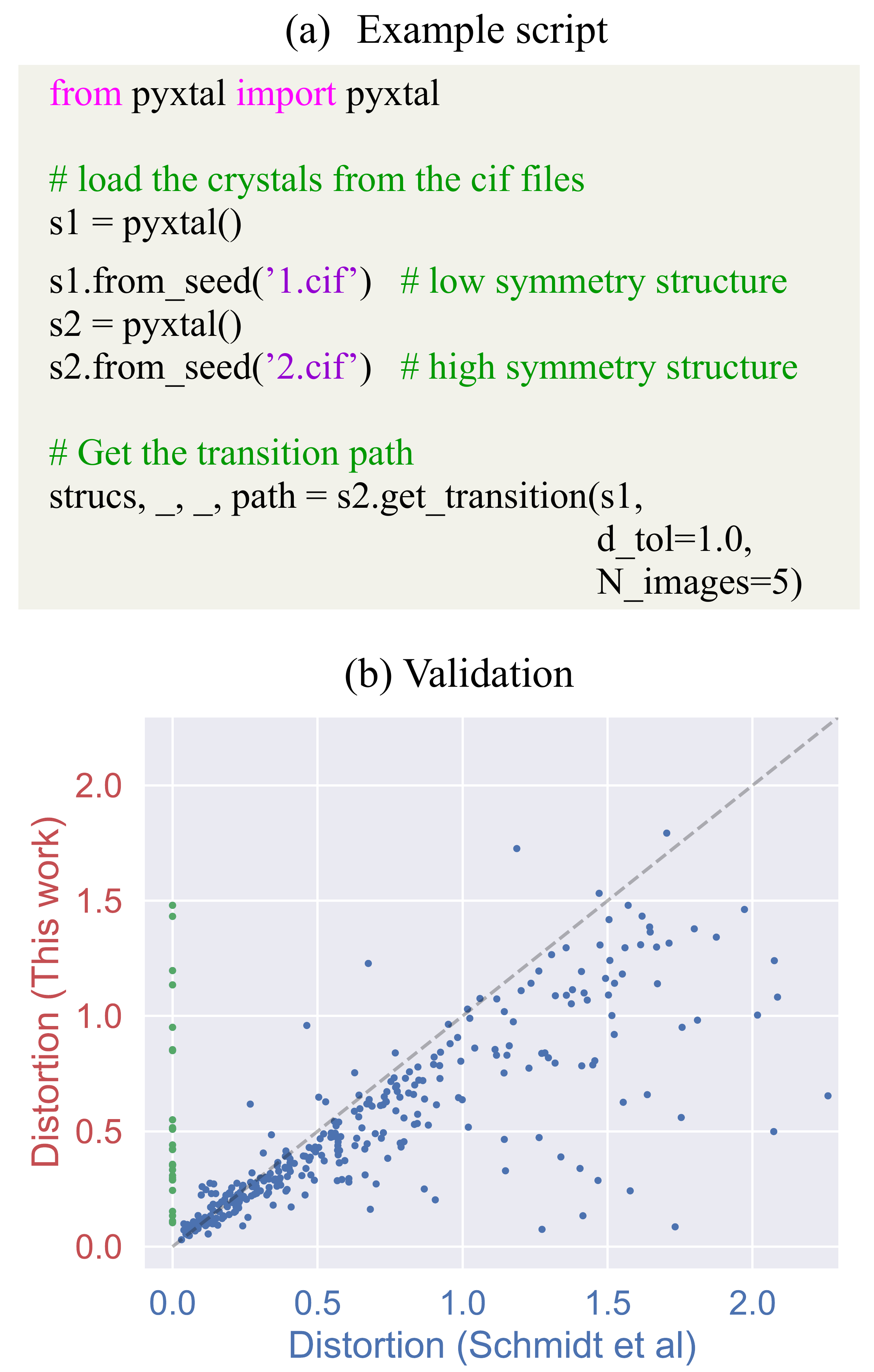}
\caption{(a) The example script to generate a transition path from two input crystals; (b) the comparison of computed maximum atomic distortion values between the current and previous work.
}
\label{fig3}
\end{figure}

This algorithm has been implemented in the open source code \texttt{PyXtal} \cite{pyxtal}. Executing the example script from Fig. \ref{fig3}a will generate a transition path (represented by 5 intermediate structures) with the smallest possible atomic distortion that satisfies the symmetry relation. 

\subsection*{2.2 Validation}
In a recent work \cite{smidt2020automatically}, Smidt and coworkers proposed 413 structure pairs that may possess continuous transformation. They also provide an online dataset with detailed information characterizing the structural, energetic and polarization quantities. To validate our algorithm implementation, we systematically checked the transition path and computed the required atomic displacements for each of the pairs. Fig. \ref{fig3}b shows the comparison of maximum distortion values between our results and the previous work. In general, our work identifies successful transition paths with smaller atomic distortions compared to those of the previous study. In that dataset, there exist 27 pairs without records for atomic distortion (shown as the green points in Fig. \ref{fig3}b), but our code can generate successful paths for these pairs with reasonably small atomic distortions. 

\section{Symmetry Relation database}\label{sec3}
With the new implementation of our automated symmetry detection algorithm, we are able to screen more materials data on a large scale. In general, researchers are interested in the phase transition involving change in polarization as it is important for a number of applications, such as ferroelectricity and piezoeletricity. Therefore, we proceed to investigate symmetry relations between polar and non-polar structure pairs with the newly developed toolkit. 

\begin{figure*}[ht]
\centering
\includegraphics[width=1.0\textwidth]{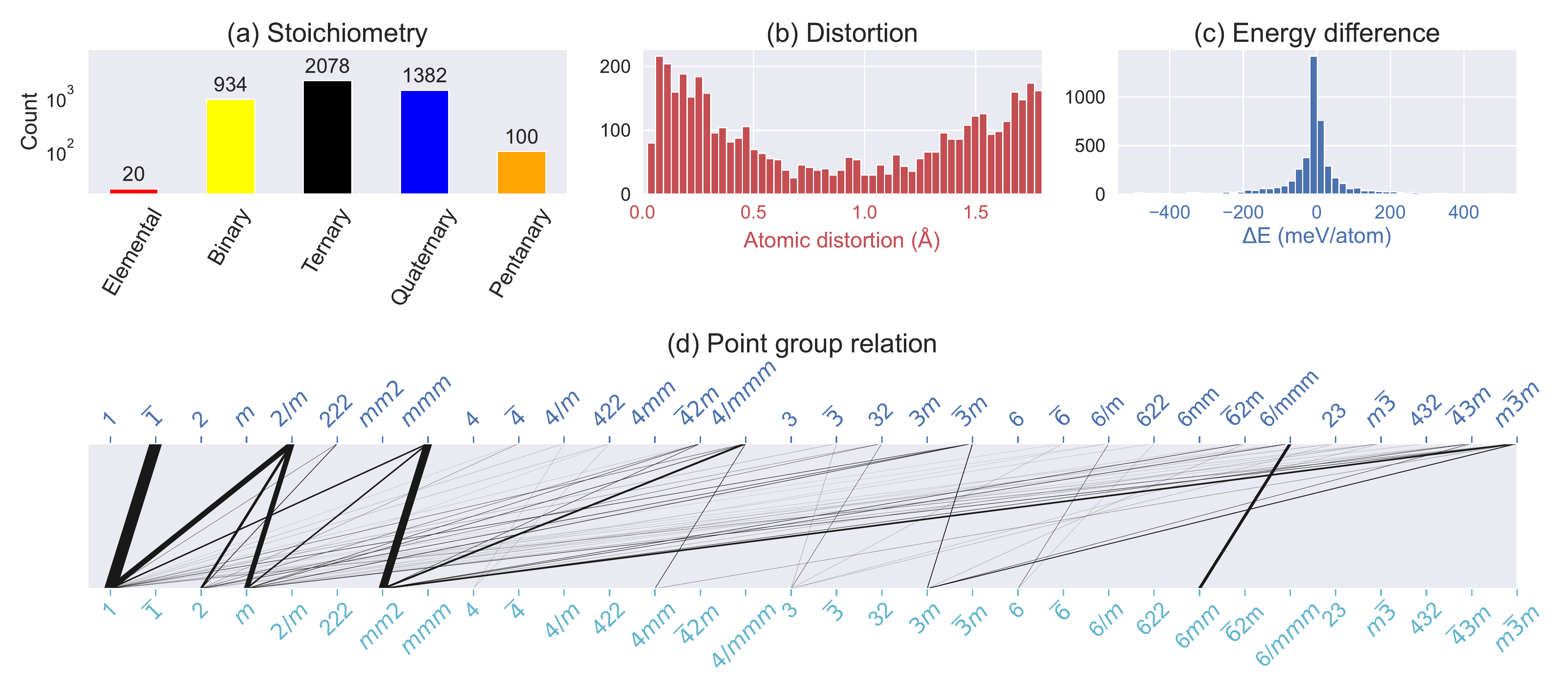}
\caption{The distribution of screened data in this work, including the distribution of (a) stoichiometry, (b) atomic distortion, (c) energy difference and (d) point group relation. In (d) the thickness of each line denotes the relative counts from the entire 4514 data.}
\label{fig4}
\end{figure*}

\subsection*{3.1 Screening of polar/non-polar structural pairs}

From the available materials data in the MP database \cite{MP-2013}, we made an initial query to obtain 28355 unique compounds belonging to polar space groups. For each polar structure, we make another query to extract the corresponding non-polar structures which may possess super group symmetry relative to the reference polar structures. We find 239101 potential nonpolar-polar symmetry pairs created from 7775 low-symmetry polar and 8051 high symmetry non-polar structures. Using the new symmetry tool, we systematically for phase transitions with a maximum atomic distortion smaller than 1.8 \AA~ and lattice mismatch smaller than 1.5 \AA~ in length and 10$^\circ$ in angle. This generates 4514 ideal structural pairs (see details in the supplementary material). 


Fig. \ref{fig4} summarizes the overall statistics. In total, we identify 20 elemental, 934 binary, 2078 ternary, 1382 quaternary and 100 pentanary polar materials. Among the 4514 structural pairs, over 55.2\% have a maximum atomic distortion smaller than 1.0 \AA. In addition, most of the structure pairs (91.7\%) have an energy difference ($\Delta E$) less than 200 meV/atom. 
Among the entire dataset, 2302 (51.0\%) pairs can be considered as high quality data to study, with phase transition within the tolerances of 1.0 \AA~ in distortion and 250 meV/atom in energy difference. Finally, Fig. \ref{fig4}d shows the frequency of point group relations. The six most frequent transitions are $\overline{1}\rightarrow1$ (21.9\%), $mmm \rightarrow mm2$ (14.6\%), $2mm \rightarrow 1$ (9.9\%), $2/m \rightarrow m$ (9.6\%), $6/mmm \rightarrow 6mm$ (5.4\%). Knowing such statistics will be useful to develop a model to predict likely continuous phase transitions for designing future new materials.

\subsection*{3.2 Ferroelectric materials identification}
From our symmetry relation database, we obtain 1234 potential ferroelectric material pairs by applying four filters: (i) the space group relation must belong to 88 potential paraelectric to ferroelectric phase transitions \cite{Aizu-PRB-1970, shi2016symmetry}; (ii) the maximum distortion is smaller than 1.0 \AA; (iii) both polar and non-polar phases have nonzero band gaps; and (iv) the energy difference between the polar and non-polar phases should be smaller than 100 meV/atom. 
Then, we systematically compute the pair's phase transition barrier and the polar structure's polarization by following the automated workflow as described in previous work \cite{smidt2020automatically} based on the VASP package\cite{kresse1996efficient}. 
Furthermore, we also choose some representative materials as candidates for further analysis. We calculate their free energy curves as a function of different external electric fields based on Landau theory\cite{chandra2007landau} and their electric hysteresis loops by finding optimized ionic positions using Hessian matrix and Born effective charges \cite{peloop1, peloop2}.

\begin{figure}[ht]
\centering
\includegraphics[width=0.48\textwidth]{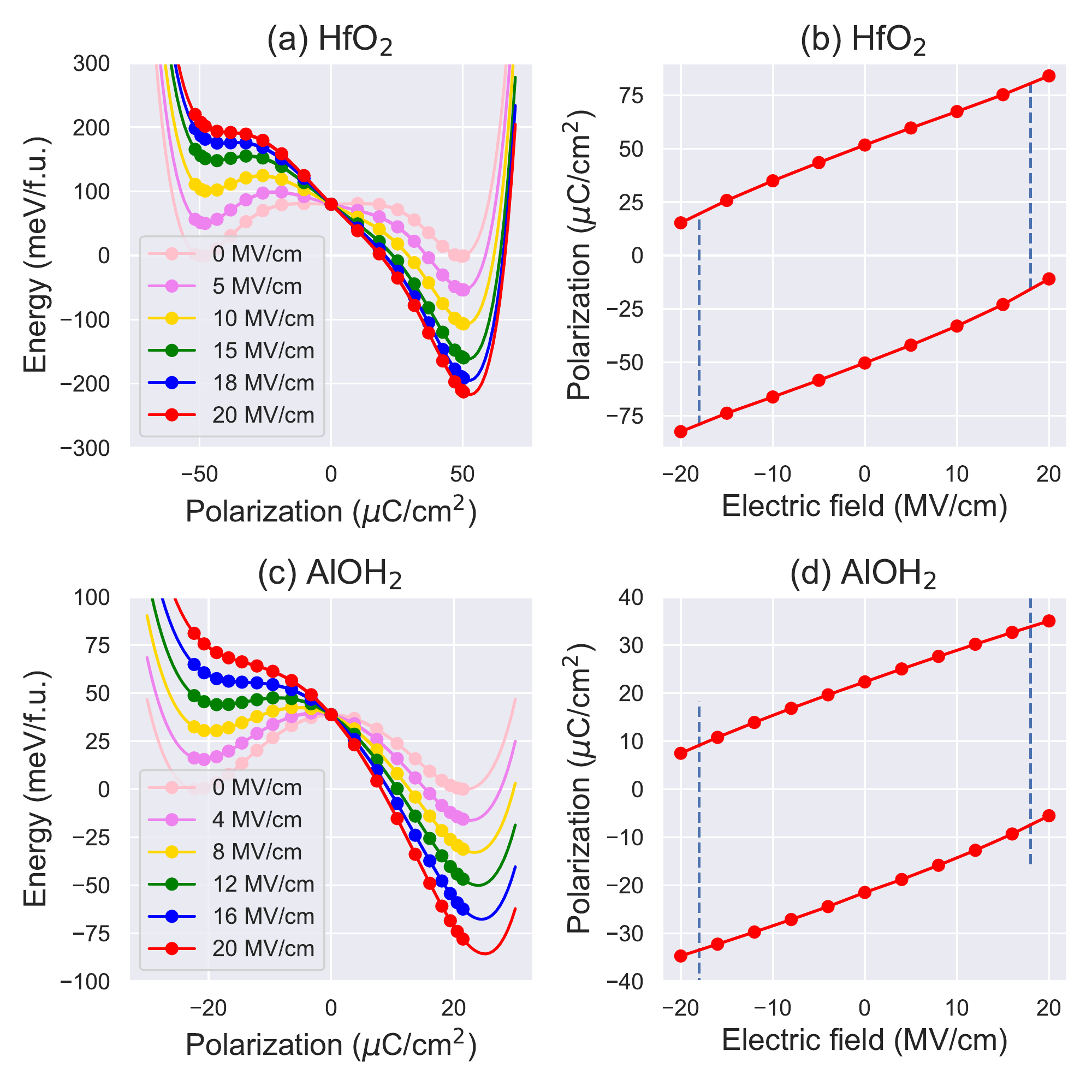}
\caption{The computational characterization of ferroelectric materials HfO$_2$ and AlHO$_2$. (a) and (c) show the free energy curves as a function of polarization for different electric fields. (b) and (d) display the corresponding computed hysteresis curves at zero temperature.}
\label{fig5}
\end{figure}

Fig. \ref{fig5} shows the results for two representative materials HfO$_2$ and AlHO$_2$ from our database. The $Pca2_1 \rightarrow P4_2/nmc$ transition of HfO$_2$ has been extensively studied in the past \cite{maeda2017identifying, fan2019origin}. Fig. \ref{fig5}a displays HfO$_2$'s potential energy curve as a function of polarization for various electric fields. At a zero electric field, the energy curve shows a typical double well behavior for ferroelectric materials, and the polar phase has a polarization of 55.2 $\mu C/cm^2$. When electric fields are applied, the free energy of the left local minimum gradually increases and the energy barrier vanishes at the coercive field (16 MV/cm). These results are consistent with previous reports based on the computational code and DFT functional \cite{maeda2017identifying}. Using this workflow, we also investigate the ferro-paraelectric phase transition of AlHO$_2$, a system that has not been considered in any previous literature. From the MP database, we find that there is a close symmetry relation between the $Pmn2_1$ (mp-23807) and $Pnnm$ (mp-1078520) phases. Compared to HfO$_2$, the polar AlHO$_2$ phase has a smaller polarization (23.1 $\mu C/cm^2$). However, the coercive field  of AlHO$_2$ (15 MV/cm) is close to that of HfO$_2$. In addition to AlHO$_2$, our screening identified 7 other materials that have not been studied in the past. These materials will be investigated in future.


\section{Conclusions}
In this work, we introduce the concept of a symmetry relation database. To enable the investigation of displacive phase transitions from historic materials data, we have implemented a numerical algorithm into the open-source code \texttt{PyXtal} that can automate the detection of symmetry relations between any two candidate crystal structures of identical chemical composition. We apply the tool to systematically screen all possible polar-nonpolar structure pairs from the entire MP database and establish a symmetry relation database of $\sim$4500 structure pairs with possible continuous phase transition paths through small atomic displacements. From this symmetry relation database, we identify several new ferroelectric materials that have never been reported in the past. 

In addition to the search of ferroelectric materials, the concept of symmetry relation can be extended to guide materials discovery in a broader sense. Most open materials databases \cite{MP-2013, Aflowlib-2012, OQMD-2015} rely the data mining techniques \cite{Hautier-IC-2010} augment available materials with hypothetical compounds constructed by decorating common structural prototypes with elements in the periodic table. In the past, chemists often found it useful to design or predict new structures by substituting one elements into two based on the isoelectronic concept (e.g., from cubic diamond type elemental C to compound BN \cite{wentorf1957cubic}, or from post-perovskite type Fe$_2$O$_3$ to MgSiO$_3$ \cite{oganov2004theoretical}). However, those works were limited to only a few case studies due to the lack of a database that can systematically describe such symmetry relations. In the future, we will extend the analysis tool to support the symmetry relation detection between different structural prototypes. Thus, one can design more powerful data mining models through more sophisticated chemical substitution.




\section*{Acknowledgments}
This research was sponsored by the U.S. Department of Energy, Office of Science, Office of Basic Energy Sciences, Theoretical Condensed Matter Physics program and the DOE Established Program to Stimulate Competitive Research under Award Number DE-SC0021970. The computing resources are provided by XSEDE (TG-DMR180040). Q.Z thanks Dr. Amois Arroyo and Dr. Shunbo Hu for inspiring discussions regarding symmetry relation and ferroelectric materials.

\section*{Data availability}
The datasets generated during and/or analysed during the current study are available at \url{https://github.com/qzhu2017/PyXtal_symmetry_relation.git}.

\section*{Conflict of interest}
All authors declare that they have no conflict of interest.

\section*{REFERENCES}
\bibliography{ref}
\end{document}